\documentclass{article}

\usepackage{PRIMEarxiv}

\usepackage[utf8]{inputenc} % allow utf-8 input
\usepackage[T1]{fontenc}    % use 8-bit T1 fonts
\usepackage[hidelinks]{hyperref}
\usepackage{url}            % simple URL typesetting
\usepackage{booktabs}       % professional-quality tables
\usepackage{amsfonts}       % blackboard math symbols
\usepackage{nicefrac}       % compact symbols for 1/2, etc.
\usepackage{microtype}      % microtypography
\usepackage{lipsum}
\usepackage{fancyhdr}       % header
\usepackage{graphicx}       % graphics
\graphicspath{{media/}}     % organize your images and other figures under media/ folder
\usepackage[numbers]{natbib}
\usepackage{amsmath}
\usepackage{bbm}

\title{Artificial Intelligence for Emergency Response
\thanks{This is a pre-print for a book chapter to appear in Vorobeychik, Yevgeniy., and Mukhopadhyay, Ayan., (Eds.). (2023). \textit{Artificial Intelligence and Society}. ACM Press.} 
}

\author{
  Ayan Mukhopadhyay \\
  Department of Computer Science \\
  Vanderbilt University \\
  Nashville, USA\\
  \texttt{ayan.mukhopadhyay@vanderbilt.edu} \\
}

\begin{document}
\maketitle

\begin{abstract}
Emergency response management (ERM) is a challenge faced by communities across the globe. First responders must respond to various incidents
, such as fires, traffic accidents, and medical emergencies. They must respond quickly to incidents to minimize the risk to human life. Consequently, considerable attention has been devoted to studying emergency incidents and response in the last several decades. In particular, data-driven models help reduce human and financial loss and improve design codes, traffic regulations, and safety measures. This tutorial paper explores four sub-problems within emergency response: incident prediction, incident detection, resource allocation, and resource dispatch. We aim to present mathematical formulations for these problems and broad frameworks for each problem. We also share open-source (synthetic) data from a large metropolitan area in the USA for future work on data-driven emergency response.

\end{abstract}

% % keywords can be removed
% \keywords{Emergency Response \and Resource Allocation \and Multi-Agent Systems}
\section{Introduction}

First responders must respond quickly to incidents such as fires, traffic accidents, and medical emergencies to minimize the risk to human life~\cite{jaldell2017important}. However, with rapid urbanization and increasing traffic volume, the losses due to emergency incidents are on the rise~\cite{mukhopadhyay2022review}, and emergency incidents still cause thousands of deaths and injuries and result in losses worth billions of dollars directly or indirectly each year~\cite{crimeUS}. Emergency management deals with proactively mobilizing resources to reduce response time and ensure that communities can recover from the effects of incidents such as accidents and fires.
Emergency response management can be divided into five major components:
1) mitigation, 2) preparedness, 3) detection, 4) response, and 5) recovery~\cite{MukhopadhyayDissertation2019,dhs,mukhopadhyay2022review}. Mitigation is the process of ensuring the long-term safety of people and property, partly by understanding when and where future incidents can occur. This process leads to the development of predictive models for risk and incident occurrence. Preparedness involves creating the necessary infrastructure for emergency response, such as optimizing the locations for responders, ambulances, and police vehicles and designing response plans. The detection process involves automated algorithms to detect incidents to improve the response speed, e.g., traffic cameras can detect accidents even before they are reported. Response involves dispatching responders when incidents happen or are about to occur. Lastly, the recovery phase ensures that impacted individuals and the community can cope with the after-effects of incidents. These stages are interlinked, with one stage often being a prerequisite to another. For example, predictive models learned in the preparedness stage are used to guide planning approaches in the response stage. These inter-dependencies must be taken into account while designing emergency management pipelines.

The design of an emergency response pipeline is often governed by the type of incident in consideration; e.g., the nature of the response differs between large-scale disasters such as hurricanes, wildfires, and floods and everyday incidents such as road accidents and crimes. In this chapter, we focus on the latter category of incidents, which enables us to explore in depth several sub-problems in this area. While such incidents are more common, they cause massive damage.  The daily death toll globally from road accidents is 3,200, which tranlsates to more than a million deaths annually~\cite{roadStats}. The sheer volume of calls that first responders receive for emergency medical services (EMS) is also staggering; more than 240 million EMS calls are made annually in the United States alone~\cite{911Stats}. First responder agencies have designed principled approaches to deal with such a large number of incidents. In the last few decades, principled data-driven approaches have been used to complement such efforts, and we will explore how artificial intelligence can be used to improve emergency response.  One way to optimize resource allocation in emergency response is to proactively  allocate resources (e.g., ambulances and fire trucks) spatially and temporally in anticipation of incidents. Indeed, considerable attention has been devoted to studying emergency response from a principled data-driven perspective~\cite{mukhopadhyayAAMAS17,erkut_ambulance_2008, knight_ambulance_2012, mccormack_simulation_2015}. 
This chapter is organized as follows. First, we will begin with a general problem formulation for emergency response, largely based on prior work by Mukhopadhyay et al.~\cite{mukhopadhyay2022review}. Then, we will explore data-driven approaches to forecasting and detecting incidents, followed by how responders can be stationed and dispatched in anticipation of future incidents. Throughout the chapter, we will use road accidents as a case study.
\section{Problem}

Emergency response enables first responders to physically travel to the spatial location of incidents such as road accidents to provide care, first aid, and transport to hospitals, among other services. While the nature of resources varies based on the type of incident and call for service, we provide an abstraction that is based on the commonalities among different incident types. The general problem structure is based on prior work by Mukhopadhyay et al.~\cite{mukhopadhyay2022review}. Given a spatial area of interest $S$, we assume that incidents occur in space and time through some \textit{true} function $f_{\text{True}}$. This function, as in other learning problems, is unknown. Instead, the decision-maker observes a set of samples (possibly noisy) drawn from an incident arrival distribution. We denote these samples, available to us in the form of a \textit{dataset}, are denoted by $D$, and can be represented as a vector of tuples
$\{(s_1,t_1,k_1,w_1),(s_2,t_2,k_2,w_2),\dots,(s_n,t_n,k_n,w_n)\}$, where $s_i$, $t_i$, and $k_i$ denote the location, time of occurrence, and reported severity of the $i$th incident, respectively, and $w_i \in \mathbbm{R}^m$ represents a vector of features associated with the incident. The vector $w$ can consist of spatial, temporal, or spatio-temporal features, capturing covariates potentially affecting the occurrence of incidents. For example, consider the occurrence of road accidents. The set $w$, in such a scenario, typically includes determinants of accidents captured by features such as weather, traffic volume, and time of day. The problem of incident prediction can then be stated as learning the parameters $\theta$ of a function over a random variable $X$ conditioned on $w$. We denote this function by $f(X \mid w,\theta)$. The random variable $X$ represents a measure of incident occurrence such as a \textit{count} of incidents (the number of incidents in $S$ during a specific time period) or \textit{time} between successive incidents. The decision-maker seeks to find the \textit{optimal} parameters $\theta^*$ that best describe $D$. This can be formulated as a maximum likelihood estimation (MLE) problem or an equivalent empirical risk minimization (ERM) problem.

The function $f(X \mid w,\theta)$, therefore, captures the likelihood (or risk) of incident occurrence given a specific set of relevant features. Given such a model, the next part of designing an emergency response pipeline is planning in anticipation of future incidents, which involves allocating (or stationing) responders strategically and dispatching them as incidents occur. This problem can be modeled by the optimization problem $\max_{y} G(y \mid f)$, where $y$ represents the decision variable (e.g., the location of emergency responders in space), $G$ is a utility function, and $f$ is the model of incident occurrence. For example, $G$ might measure the total coverage (spatial spread) of the responders or the expected response time to incidents. Therefore, given a model of incident occurrence $f$, the decision-maker's goal is to maximize the utility defined by $G$. The optimization can be represented as a sequential stochastic control process and modeled by a Markov decision process (MDP)~\cite{kochenderfer2015decision, mukhopadhyayAAMAS18}. Such a formulation can tackle the temporal nature of the response and is particularly relevant for finding a general mapping (i.e., a policy) for response stationing and dispatch. The second approach is to directly model the allocation problem as a single-shot optimization problem according to a specific measure of interest, e.g., one way to model resource stationing is maximizing the coverage of emergency responders~\citep{toregas_location_1971,church1974maximal, gendreau_solving_1997, mukhopadhyayAAMAS17}.

Emergency response pipelines were traditionally designed in a manner that initiated response \textit{after} an incident was reported, e.g., through 911 calls in the USA. However, in practice, there is a gap in time between the occurrence of an incident and the time when it is reported. This loss in time can be detrimental to the response, especially in life-threatening incidents such as accidents. This gap can be bridged by using incident detection or extraction models, which use crowdsourced and sensor data to detect incidents automatically. For example, traffic cameras can be used to detect accidents as soon as they occur. Mathematically, this process involves learning a function $E$ which takes relevant data $u$ as input (for example, video data from a traffic camera) and outputs information about an event (for example, the location of an accident). This can be represented as $\bar{z} = E(u)$, where $\bar{z}$ denotes an estimate of variable $z$. In event extraction, $z$ can denote any variable related to the incident, such as the location of the incident, the extent of the damage, the number of people involved, and so forth.
\section{Forecasting Incidents}\label{sec:prediction}

The most important aspect of using data-driven approaches for emergency response is to understand \textit{where} and \textit{when} are incidents likely to occur. There is a plethora of prior work in this domain, ranging from using the historical frequency of incidents to learn an empirical distribution~\cite{deacon1974identification} to using neural networks that use heterogeneous features~\cite{vazirizade2021learning}. Early approaches often used linear regression models~\cite{frantzeskakis1994interurban,jovanis1986modeling} to predict the number of likely incidents in an area. However, such models fail to capture the discrete and strictly positive nature of incident counts. Moreover, they also fail to account for the sporadic nature of emergency incidents~\cite{miaou1993modeling,joshua1990estimating}. An alternate approach is to use Poisson models, which have been used widely to forecast incidents such as road accidents~\cite{jovanis1986modeling,lord2005poisson,bonneson1993estimation,maher1996comprehensive,sayed1999accident,joshua1990estimating}. Poisson models can also suffer from two disadvantages in this context: first, they assume that the expected value of the response variable (count of incidents) is equal to its variance. This assumption does not usually hold in practice with emergency data~\cite{lord2005poisson,mukhopadhyay2022review}. Second, such models fail to adequately take into account the prevalence of zero counts in real-world data pertaining to emergency incidents. Consider data collected for road accidents at an hourly frequency; most roads do not experience accidents every hour, and as a result, such data often has very high sparsity~\cite{vazirizade2021learning,mukhopadhyay2022review,lord2005poisson}. To tackle these problems, zero-inflated Poisson models and zero-inflated negative binomial models are often used~\cite{mukhopadhyay2022review}. Another approach to account for high sparsity in data is to use resampling-based approaches. For example, it is possible to either over-sample the positive labels (i.e., the occurrence of incidents) or under-sample the lack of incidents to balance a dataset~\cite{vazirizade2021learning}.

The availability of big data has resulted in the creation of richer feature sets, which in turn, has enabled the usage of more complex algorithmic models for incident forecasting. For example, neural networks and ensemble learning have been extensively used in the context of forecasting road accidents~\cite{Abdel-Aty2008,Yu2014,Pande2006,Abdelwahab2002,Chang2005,Riviere2006,zhu2018use,Bao2019}. An important consideration of developing complex models is data collection and feature engineering. Emergency incidents are caused by a variety of factors; it is recommended that model designers collaborate with domain experts to identify the determinants of specific types of incidents. However, such data is often available at different temporal and spatial scales. For example, while traffic data can be mined in (almost) real-time, demographic data is typically available over much larger time intervals. Similarly, traffic data can be mapped to a graph (e.g., the existing road network), whereas demographic data is usually available over a grid or census tracts. Therefore, model designers often need to harmonize the heterogeneous data into the same spatial and temporal scale before learning models.
Broadly, features for incident prediction (especially in the context of roadway accidents) can be divided into temporal, spatial, or a combination of both~\cite{mukhopadhyay2022review}. For example, a commonly used feature in incident prediction approaches is the time of day or the season (e.g., summer or winter), which are examples of temporal features. The geometry of a specific road segment, on the other hand, is a spatial feature, as it is a property of a specific spatial unit. Spatio-temporal features measure spatial properties that change with time, such as traffic congestion on a particular road segment. 

Finally, in addition to using a rich variety of features, model designers must account for additional checks to ensure that models accurately represent the underlying process of incident occurrence. For example, unobserved heterogeneity can arise in a model when features that are not recorded in the data affect the occurrence of incidents or are correlated to observed features~\cite{mannering2014analytic,mukhopadhyay2022review}. Similarly, data collection can be biased for many reasons. For example, in the example of roadway accidents, vehicle crashes with no severe injuries are often under-reported, thereby causing sampling bias~\cite{yasmin2014latent}. Such biases can be particularly detrimental when data-driven approaches are used for law enforcement, where existing enforcement practices can produce biased data against specific communities or spatial areas.

\section{Detecting Incidents}

Emergency response to an event can only be dispatched after the concerned agency (e.g., the fire department) is notified about the event. Traditionally, the pipeline for emergency response relied on manual notification, i.e., either the victim of the event or a passer-by would call a helpline number (e.g., 911) to request assistance. However, the delay between when an event actually occurs and when assistance is requested can be critical in practice. For example, consider a road accident where the victim is severely injured and unable to call for assistance; waiting for a passer-by to call 911 could be life-threatening for the victim. However, modern cities are now equipped with various modalities of data streams and sensors that can be used to detect such events quickly (even before they are reported). For example, consider a fire in an urban area. Residents who observe the fire can report it on social media (even before they call an official helpline). Similarly, road accidents can be detected by traffic cameras. The goal of designing incident detection pipelines in emergency response is to use heterogeneous data to detect the occurrence of events in order to reduce the overall time for response.

While event detection can focus on various modalities of data, the three most commonly used (and promising) modalities are the following: using text data from social media (e.g., from applications such as Twitter), video or image data (e.g., data from traffic cameras), and crowdsourced event reports (e.g., from applications such as Waze). In general, there are two subtasks in event detection for emergency response~\cite{mukhopadhyay2022review}: 1) the first task seeks to identify the occurrence of the event and relies on ``triggers'' that aid detection; e.g., a social media post containing words such as \textit{accident} or \textit{crash} could be used as triggers. Naturally, complex models leverage more abstract patterns to detect the occurrence of incidents. 2) The second task deals with identifying information about the incident, e.g., where the incident has happened, what is the most likely time of occurrence, how many people (or structures) are involved, and so on. Natural Language Processing (NLP) techniques have widely been used to extract event information from social media text data. One of the traditional approaches to tackle this problem is to use feature engineering~\cite{grishman2005nyu,ahn2006stages,ji2008refining}. First, the input data (text) is converted into a sequence of tokens. Then, for each token, various features such as lexical and syntactic features are extracted. These features are then used to learn a classifier. One limitation of this approach is that social media text is often unstructured and has informal grammar. Recently, deep neural networks have become an attractive choice as no manual feature engineering is required~\cite{chen2015event,nguyen2016joint}.

Event extraction is also possible from video and image data, especially from traffic cameras, which can be used to track and monitor congestion and detect incidents that need response~\citep{gan2015devnet,chang2016bi,ma2017joint,li2020cross}. For example, Gan et al.~\citep{gan2015devnet} use a CNN, pre-trained on a large corpus of openly available image data to perform keyframe detection. More recently, there has been work that proposes a new approach that projects the structured semantic representations from the combination of textual and visual data into a common space, which then aids learning~\citep{li2020cross}.
Finally, Event detection from crowdsourced event reporting data has also garnered interest in recent years, and several first responder agencies have started using applications such as Waze~\cite{waze1, waze2}. Senarath~\cite{yasas2021} leverage crowdsourced data to detect roadway accidents and point out that the accuracy of detection has a trade-off with spatial and temporal localization. Moreover, it is imperative to consider that practitioners might focus on different metrics for effective deployment (e.g., reduced number of false alerts and a balance between precision and recall) than standard metrics such as accuracy.
\section{Responding to Incidents}

Responding to incidents is the most consumer-facing component of the emergency responder pipeline. However, a critical prerequisite of  response is resource allocation, i.e., optimizing resources spatially and temporally in anticipation of incidents. This anticipation, or forecast, is fed to the allocation in the form of a prediction model, which we discussed previously in section~\ref{sec:prediction}. As pointed out by Mukhopadhyay et al.~\cite{mukhopadhyay2022review}, the distinction between allocation and response can be unclear as a solution to the allocation problem can implicitly define a response strategy. For example, consider an allocation problem, whose solution minimizes the expected response times to incidents defined by a distribution $f$. Responders such as fire trucks or ambulances can be stationed based on this solution. Now, when an incident occurs, the closest available responder might be dispatched, which is a response strategy often followed in practice by first responders. However, there are several nuances involved that must be considered, e.g., dispatching the closest responder does not necessarily minimize response times in the long run and the severity of the incident might dictate which responder to dispatch. The problem of responder dispatch seeks to address such questions in a principled manner.

The allocation problem is often formulated as a \textit{single-shot} combinatorial optimization problem that seeks to maximize metrics such as the spatial coverage of the allocated responders~\cite{toregas_location_1971,church1974maximal} or the probability of patient survival~\cite{erkut_ambulance_2008, knight_ambulance_2012}. The two canonical formulations that maximize coverage are the Location Set Covering Problem (LSCP) \citep{toregas_location_1971} and the Maximal Covering Location Problem (MCLP) \citep{church1974maximal}; the former seeks to find the least number of facilities (e.g., fire stations) that \textit{cover} all spatial areas that need service (e.g., based on municipal jurisdiction), the latter maximizes the overall area covered by a given number of facilities. It is also common to add secondary goals as constraints to the coverage problem, such as an upper bound on the maximum time an incident must wait before a responder arrives on the scene~\cite{silva2008locating, mukhopadhyayAAMAS17}. Two major shortcomings of using the facility location problem to allocate emergency responders are the strong assumptions that responders are always available at their locations and a facility is able to account for the entire area that it \textit{covers}. To tackle this challenge, several extensions have been proposed, such as the Double Standard Model (DSM)~\citep{gendreau_solving_1997}, which posits that while every area must have \textit{some} coverage, some areas, e.g., the ones with the majority of incidents, have greater coverage. There is also extensive work on probabilistic coverage models, which model the stochastic nature of station availability~\cite{daskin_maximum_1983, revelle_maximum_1989}. 

While allocation problems are typically formulated as integer linear programs, the response problem presents a sequential decision-making problem under uncertainty---as incidents happen over time, responders must be dispatched and their distribution must be re-adjusted given the uncertainty over the occurrence of future incidents. This problem is typically modeled as a Markov decision process (MDP), which is a natural model for discrete-time stochastic control processes~\cite{kochenderfer2015decision}. The decision-maker's goal is to choose an \textit{action} (e.g., which responder to dispatch) given a \textit{state}; the state subsumes relevant information of interest, e.g., how many responders are available, what is the expected demand for responders, and so on. The MDP is often solved by learning an optimal policy, which is a mapping from states to actions, that maximizes expected utility or minimizes expected cost. The cost function can be constructed based on domain expertise, and in its simplest version could simply be the time taken for response.

Early work on using MDPs to tackle emergency response leverage continuous-time MDPs, which assume that the time between the world transitioning from one state to another is memoryless~\cite{carter1972response, keneally2016markov}. While this framework provides computational convenience, real-world problems violate these assumptions, e.g., the time taken for an ambulance to travel to the scene of the incident is not memoryless. Mukhopadhyay et al.~\cite{mukhopadhyayAAMAS18} relax this assumption by considering a semi-Markovian decision process. While this framework is more challenging due to absence of closed-form transition functions for the MDP, it is more suitable to emergency response and historical data can be used to learn the transition distribution. Often times, learning a policy for the MDP can be intractable due to the large state and action space that emergency response poses. In such situations, decision-makers can leverage online planning, where instead of learning a general mapping from all states to actions, the goal is to quickly compute a near-optimal action for the current state of the world~\cite{MukhopadhyayICCPS}. Other approaches to tackle the large state and action space are to do decentralized~\cite{pettet2020algorithmic} or hierarchical planning~\cite{pettet2020hierarchical}, which segregate the planning problem into smaller sub-problems that are solved independently. For example, one way to perform hierarchical planning might be to identify spatial regions (similar to jurisdictions) within a city and solve an independent MDP for each region.

An important consideration in designing response strategies is the importance of a \textit{greedy dispatch} strategy depending on the severity of some categories of incidents. In practice, our conversations with first responders revealed that the ethical and legal constraints under which first responders operate could dictate that the closest available responder be dispatched to critical incidents, even when such a dispatch strategy is proven to be sub-optimal in the long run. Since the exact severity of an incident is difficult to gauge in practice, first responders often leverage a greedy dispatch strategy for a wide array of incidents. To ensure that the overall expected response time does not suffer too much, Pettet et al.~\cite{pettet2020hierarchical} propose dynamic reallocation with greedy dispatch; in such an approach, when an incident occurs and a call for emergency service is required, the closest responder is always dispatched. However, the allocation of responders is periodically re-adjusted to account for temporal changes or the unavailability of responders in some regions. Dynamic reallocation has been shown to improve response times in emergency response simulations.

\section{Conclusion}

Emergency response is a critical piece of infrastructure required around the globe. First responders face the daunting challenge of responding to an increasing number of calls for service with limited budgets and time. Artificial intelligence and data-driven decision-making can learn complex abstractions from historical data to enable first responders to make better decisions in practice. While automated decision-making cannot (at least yet) replace the invaluable domain expertise of first responders, principled automated approaches for prediction, detection, and response are being adopted by cities across the world to expedite and improve emergency response.

%Bibliography
\bibliographystyle{unsrt}  
\bibliography{references}

\end{document}